\documentclass{ifacconf}

\usepackage{setspace}
\usepackage{url}
\usepackage[utf8]{inputenc}
\usepackage[T1]{fontenc}
\usepackage{graphicx}		
\usepackage{wrapfig}
\usepackage[format=plain,font=footnotesize,labelfont=bf,labelsep=period]{caption}
\usepackage{sidecap} 
\usepackage{subfig}
\usepackage[export]{adjustbox}
\usepackage[font=small]{caption}
\usepackage{float}
\usepackage[table]{xcolor} 

\usepackage{amsmath} 
\allowdisplaybreaks
\usepackage{amssymb}  
\usepackage{mathtools}
\usepackage[normalem]{ulem}
\usepackage{paralist}	
\usepackage[space]{grffile} 
\usepackage{color}

\usepackage{enumitem}
\usepackage{bm}
\usepackage{cancel}
\usepackage{hhline}
\usepackage{multirow}

\usepackage{etoolbox} 

\newcommand{\uu}{\mathbf{u}}
\newcommand{\UU}{\mathbf{U}}
\newcommand{\bbp}{\mathbb{P}}
\newcommand{\bbe}{\mathbb{E}}

\newcommand{\Xcal}{\mathcal{X}}

\definecolor{darkblue}{RGB}{0,0,102}
\definecolor{lightblue}{RGB}{77,77,148}

\definecolor{gold}{RGB}{234, 170, 0}
\definecolor{metallic_gold}{RGB}{139, 111, 78}

\newtoggle{ShowRemarks}
\togglefalse{ShowRemarks} 
\iftoggle{ShowRemarks}{
\newcommand{\TODO}[1]{\textcolor{red}{{\bf TODO:}  #1}}
\newcommand{\NOTE}[1]{\textcolor{red}{{\bf NOTE:}  #1}}
\newcommand{\NL}[1]{\textcolor{red}{{\bf NL:}  #1}}
\newcommand{\CRH}[1]{\textcolor{blue}{{\bf CRH:}  #1}}
\newcommand{\UPDATE}[1]{\textcolor{blue}{#1}}
}{
\newcommand{\TODO}[1]{}
\newcommand{\NOTE}[1]{}
\newcommand{\NL}[1]{}
\newcommand{\CRH}[1]{}
\newcommand{\UPDATE}[1]{#1}
}

\usepackage{xfrac}

\usepackage{dblfloatfix}

\usepackage{comment}

\usepackage{natbib}

\begin{document}

\begin{frontmatter}
\title{
Planning Persuasive Trajectories Based on \\ a Leader-Follower Game~Model
}

\author{Chaozhe R. He$^{1}$, Yichen Dong$^{2}$, Nan Li$^{2}$}
    
\thanks[footnoteinfo]{$^{1}$C. R. He is with the Department of Mechanical and Aerospace Engineering, University at Buffalo, Buffalo, NY, US. ${\tt\small chaozheh@buffalo.edu}$}%
\thanks[footnoteinfo]{$^{2}$Y. Dong and N. Li are with the School of Automotive Studies, Tongji University, Shanghai, China. ${\tt\small dych.happy@163.com}$, ${\tt\small li\_nan@tongji.edu.cn}$}%

\pagestyle{plain}           

\begin{abstract}
We propose a framework that enables autonomous vehicles (AVs) to proactively shape the intentions and behaviors of interacting human drivers. 
The framework employs a leader-follower game model with an adaptive role mechanism to predict human interaction intentions and behaviors. 
It then utilizes a branch model predictive control (MPC) algorithm to plan the AV trajectory, persuading the human to adopt the desired intention. 
The proposed framework is demonstrated in an intersection scenario. Simulation results illustrate the effectiveness of the framework for generating persuasive AV trajectories despite uncertainties. 
\end{abstract}

\end{frontmatter}

\section{Introduction}\label{sec:intro}
\vspace{-1.0mm}

Autonomous vehicles (AVs) have the potential to transform the landscape of ground transportation by enhancing safety, increasing efficiency, and expanding accessibility. 
However, successfully integrating AVs into existing road systems requires them to handle complex interactions with other road users, particularly human-driven vehicles (HVs) \cite{schwarting2019social,wang2022social}. 
A key source of complexity stems from the variability in human driver intentions and corresponding behaviors (e.g., proceeding or yielding). 
Various factors, including personal preferences, emotional states, and the driving context, influence driver intentions \cite{xing2019driver}. 
In traffic scenarios that involve extensive interactions (e.g., negotiation, competition, and cooperation), such as at unsignalized intersections, highway on/off-ramps, or in congested traffic, AVs may not only respond to human driver behaviors but also proactively shape their intentions to promote cooperation. 
This paper aims to develop an autonomous driving algorithm that can safely respond and further effectively shape the intention and behavior of human drivers, thereby improving AVs' ability to navigate complex interaction scenarios.

\subsubsection{Related Work}
Game theory provides a mathematical framework for studying the interactions of intelligent agents. 
Various researchers have utilized it in developing algorithms for modeling interaction traffic scenarios and predicting vehicle behaviors in these scenarios, e.g., in \cite{bahram2015game,yu2018human,zhang2019game,li2017game,tian2020game,li2020game,liu2022interaction,he2024ecodriving}.

To account for variable human driver intentions, \cite{zhang2019game} incorporates the driver's aggressiveness as a model parameter and estimates it online based on observed vehicle behavior; \cite{li2017game} and \cite{tian2020game} model human drivers of different types based on cognitive levels and different driver types can have different intentions under the same/similar traffic condition; \cite{lefkopoulos2020interaction} uses multiple models to predict vehicle behaviors under different driver intentions and uses a hybrid Kalman filtering scheme to identify the most likely model.

In \cite{li2020game}, we presented a leader-follower game model (LFG, extended from the classical Stackelberg game model) for modeling the interactions of multiple vehicles. 
In this model, different driver intentions in an interaction traffic scenario, particularly the intention to proceed ahead or yield to other vehicles, are represented by the asymmetric leader and follower roles of the game players. 
In \cite{liu2022interaction} and \cite{he2024ecodriving}, we utilized this leader-follower game model to develop algorithms for AVs to navigate traffic scenarios involving extensive interactions with HVs safely. 
\UPDATE{LFG provides~a~lightweight conflict resolving mechanism when interacting with~humans.}

However, the above approaches can, at most, estimate the driver's intentions based on observed past behaviors of the interacting vehicles and then optimally respond to the estimated intentions -- none of these approaches can proactively shape the intentions of human drivers. 
In light of the benefits of promoting human-robot cooperation, methods for enabling robots to proactively shape human intentions and behaviors were proposed in \cite{parekh2022rili} and~\cite{pandya2024towards}. 
However, human-robot interactions are not represented by a game model in the methods of \cite{parekh2022rili} and \cite{pandya2024towards}, and the proposed methods may not be readily applicable to autonomous driving applications.

\textit{Main Contributions} \quad 
In this work, we propose a novel framework for autonomous agents to proactively influence the intentions and behaviors of interacting humans using game theory. 
The framework is particularly well-suited for autonomous driving applications, promoting AV-HV cooperation and enhancing both safety and efficiency. 
\begin{itemize}
    \item We extend the leader-follower game model of \cite{li2020game,liu2022interaction,he2024ecodriving} to one with a role-adapting mechanism and utilize it for predicting human interactive intention and behavior. 
    \item We design a branch MPC algorithm that can predict HV role changes in response to AV behaviors and utilize it for planning persuasive AV trajectories.
    \item We demonstrate, through a simulated intersection scenario, that the proposed planning framework can generate proactive AV behavior that influences human driver intention without compromising safety.
\end{itemize}

The remainder of this paper is organized as follows. First, we formulate the problem in Section \ref{sec:background}. We present the leader-follower game model with an adaptive role in Section \ref{sec:LFG} and the interactive planning algorithm in Section \ref{sec:interactive_planning}. 
A case study on autonomous driving at the intersection is given in Section \ref{sec:case_study} to showcase their effectiveness. 
We conclude the paper \ref{sec:conclusion} and discuss future directions.

\vspace{-1.0mm}
\section{Problem Statement}\label{sec:background}
\vspace{-1.0mm}

In this section, we present the AV-HV interaction problem of interest as a general human-robot interaction problem. In particular, we consider the following 2-player human-robot interaction game: 
\begin{equation}\label{eqn:compact_dynamics_openloop}
    x_{t+1} = f(x_{t}, u^{\rm H}_{t}, u^{\rm R}_{t})
\end{equation}
where $x = (x^{\rm H}, x^{\rm R})$, the ``H'' stands for ``human'' and the ``R'' stands for ``robot.'' We assume the robot has a reward function that encodes its goal:
\begin{equation}\label{eqn:step_reward_generic}
    R(x, u^{\rm R}, u^{\rm H})
\end{equation}
which depends on the states and actions of both players. 
With this reward function, the robot adopts an MPC-based control strategy: At each time $t$, the robot computes an optimal action sequence $\uu^{\rm R, \ast}_{t} = \{u_{t}^{\rm R, \ast}, u_{t+1}^{\rm R, \ast},\ldots, u_{t+N-1}^{\rm R, \ast}\}$ that maximizes the cumulative reward over the planning horizon, i.e.,
\begin{equation}\label{eqn:mpc_policy}
    \begin{aligned}
    \uu^{\rm R, \ast}_{t} \in \,&\, \underset{\uu^{\rm R}_{t}}{\arg\max}
    \sum_{\tau = 0} ^{N-1} \lambda^{\tau} R\left(x_{t + \tau + 1}, u^{\rm R}_{t+\tau}, \hat{u}^{\rm H}_{t+\tau}\right) \\
    \text{s.t.} &\,\, x_{t+\tau + 1} = f(x_{t+\tau}, \hat{u}^{\rm H}_{t+\tau}, u^{\rm R}_{t+\tau}) \\
    &\,\, x_{t+\tau} \in \Xcal_{\rm safe},\quad \forall \tau = 1,\ldots, N
    \end{aligned} 
\end{equation}
where $\lambda \in (0,1]$ is a discount factor, and $\Xcal_{\rm safe}$ represents a safe state set used to represent strict safety specifications.

A decision process as above requires prediction of human actions, i.e., the $\hat{u}^{\rm H}_{t+\tau}$ in \eqref{eqn:mpc_policy}, where ``$\wedge$'' is used to distinguish predicted actions from the human player's actual actions. Traditional methods typically utilize a ``predict then plan'' strategy, where $\hat{u}^{\rm H}_{t+\tau}$ are first determined and remain fixed as the robot plans the actions $u^{\rm R}_{t+\tau}$. 
Such a strategy is lightweight in computation but misses the action-reaction interactive nature of the game during planning. To develop a computationally tractable algorithm that enables interactive planning, in particular, proactively shaping predicted human actions $\hat{u}^{\rm H}_{t+\tau}$ during planning $u^{\rm R}_{t+\tau}$, we leverage a leader-follower game model with a role-adapting mechanism, introduced in the next section.

\vspace{-1.0mm}
\UPDATE{\section{Adaptive Leader-Follower Game}}\label{sec:LFG}
\vspace{-1.0mm}

In this section, we first briefly review the leader-follower game (LFG) model of \cite{li2020game,liu2022interaction,he2024ecodriving} and then present the proposed role-adapting mechanism. 
The extended LFG model with the role-adapting mechanism will be utilized to predict human interactive intentions and behaviors.

\vspace{-1.0mm}
\subsection{Leader-Follower Game}\label{sec:LFG_1}
\vspace{-2.0mm}
Consider a pair of players, one is a ``leader'' and the other is a ``follower'', and rewrite the cumulative reward in \eqref{eqn:mpc_policy}~as
\begin{equation}
\small
\bar{R}_{\sigma}\left(x_t, \uu_{{\rm l}, t}, \uu_{{\rm f}, t}\right) = 
\sum_{\tau = 0}^{N-1} \lambda^{\tau}R_{\sigma}\left(x_{t+\tau + 1}, u_{{\rm l}, t+\tau}, u_{{\rm f}, t+\tau}\right)
\end{equation}
where the subscript $\sigma \in \{ \rm leader (l), follower (f)\}$ indicates the role of the player in the game; $\uu_{{\rm l}, t} = \{u_{{\rm l}, t+\tau}\}_{\tau = 0}^{N-1} \in \UU_{\rm l}$ and $\uu_{{\rm f}, t} = \{u_{{\rm f}, t+\tau}\}_{\tau = 0}^{N-1} \in \UU_{\rm f}$ denote the action sequences of the leader and the follower; $\UU_{\rm l}$ and $\UU_{\rm f}$ are the admissible sets of action sequences of both roles; and $R_{\sigma}(\cdot, \cdot, \cdot)$ is a single-step reward similar to that defined in \eqref{eqn:step_reward_generic}, where the subscript $\sigma$ implies that the reward functions for the two different roles can be different.

The leader and the follower both aim to maximize their cumulative rewards but follow different strategies: The follower maximizes the worst-case reward due to uncertain leader's actions, i.e., it takes the following ``max-min'' strategy:
\begin{subequations}\label{eqn:follower_decision_strategy}
\begin{equation}
\uu^{\ast}_{{\rm f}, t} \in \underset{\uu_{\rm f} \in \UU_{\rm f}}{\arg\max}\, Q_{\rm f}(x_t, \uu_{\rm f})
\end{equation}
where
\begin{equation}
Q_{\rm f}(x_t, \uu_{\rm f}) = \min_{\uu_{\rm l} \in \UU_{\rm l}} \bar{R}_{\rm f}(x_t, \uu_{\rm l}, \uu_{\rm f}).
\end{equation}
\end{subequations}
This strategy represents an intention to yield to the leader~\cite{li2020game}. In contrast, the leader assumes that the other player is a follower who uses the above ``max-min'' strategy. Hence, the leader can predict the follower's actions and optimize its actions according to:
\begin{subequations}\label{eqn:leader_decision_strategy}
\begin{equation}
\uu^{\ast}_{{\rm l}, t} \in \underset{\uu_{\rm l} \in \UU_{\rm l}}{\arg\max}\, Q_{\rm l}(x_t, \uu_{\rm l})
\end{equation}
where
\begin{align}
& Q_{\rm l}(x_t, \uu_{\rm l}) = \min_{\uu_{\rm f} \in \UU_{\rm f}^{\ast}(x_t)} \bar{R}_{\rm l}(x_t, \uu_{\rm l}, \uu_{\rm f}) \\
& \UU_{\rm f}^{\ast}(x_t) = \{ \uu_{\rm f} \in \UU_{\rm f}: Q_{\rm f}(x_t, \uu_{\rm f}) \geq Q_{\rm f}(x_t, \uu_{\rm f}'), \forall \uu_{\rm f}' \in \UU_{\rm f}\}.
\end{align}
\end{subequations}
This strategy represents a player that assumes the other player will yield and hence decides to proceed ahead~\cite{li2020game}. It is shown in \cite{li2020game,liu2022interaction,he2024ecodriving} that the above LFG model can accurately predict vehicle interactive behaviors in various traffic scenarios, including intersections \cite{li2020game} and highways \cite{liu2022interaction,he2024ecodriving}. We note that in the considered human-robot interaction game, either player, $\bullet \in \{{\rm H, R}\}$, can be a leader or follower, and they may not hold complementary roles (i.e., they may both be leaders or followers). For more comprehensive discussions on the above LFG model, please refer to \cite{li2020game} or \cite{liu2022interaction}.

\vspace{-2.0mm}
\subsection{Leader-Follower Game with Adaptive Roles}\label{sec:LFG_2}
\vspace{-2.0mm}

In the LFG model, the role $\sigma \in \{\rm l, f\}$ represents the player's intention and is a key parameter determining the player's behavior. 
It is assumed to be constant over the interaction period in \cite{li2020game} and \cite{liu2022interaction}. 
In this work, we relax this assumption and allow $\sigma$ to change over time, motivated by the observation that a human driver can dynamically adjust their intention in response to changing traffic conditions \UPDATE{\cite{tian2021learning}}.

Let $\bullet$ denote the ego player and $\circ$ denote the other player. 
We assume that the ego player holds a belief on the constant role of the other player, $\sigma^{\circ}$, and updates the belief based on Bayesian inference as follows \cite{hwang2006state}:
\begin{equation}\label{eqn:role_prop_update}
    \bbp(\sigma^{\circ} = \varepsilon | \xi_t) \propto \bbp(x_t | \sigma^{\circ} = \varepsilon) \bbp(\sigma^{\circ} = \varepsilon | \xi_{t-1})
\end{equation}
where $\propto$ indicates ``proportional to''; $\xi_t$ denotes the collection of observed information up to time $t$, i.e.,
\begin{equation}
    \xi_t = \{x_0, \ldots, x_{t-1}, x_t, u^{\bullet}_0, \ldots, u^{\bullet}_{t-1}\}
\end{equation}
where $x_0, \ldots, x_t$ are observed game states and $u^{\bullet}_0, \ldots, u^{\bullet}_{t-1}$ are the ego player's own actions taken at previous times; and the ``likelihood function'' $\bbp(x_t | \sigma^{\circ} = \varepsilon)$ is given as
\begin{equation}\label{eqn:observation_likelihood}
    \bbp(x_t | \sigma^{\circ} = \varepsilon) = \mathcal{N}(r_t|_{\sigma^{\circ} = \varepsilon}, 0, \mathcal{W}).
\end{equation}
In \eqref{eqn:observation_likelihood}, $\mathcal{N}(\cdot, 0, \mathcal{W})$ denotes the probability density function of the multivariate normal distribution with zero mean and covariance $\mathcal{W}$ evaluated at $(\cdot)$; $r_t|_{\sigma^{\circ} = \varepsilon}$ is the residual between observed state and predicted state assuming the other player's role is $\varepsilon$, i.e.,
\begin{equation}\label{eqn:prediction_residue}
    r_t|_{\sigma^{\circ} = \varepsilon} = x_t - \hat{x}_t|_{\sigma^{\circ} = \varepsilon}.
\end{equation}
The predicted state $\hat{x}_t|_{\sigma^{\circ} = \varepsilon}$ is given by
\begin{equation}
    \hat{x}_t|_{\sigma^{\circ} = \varepsilon} = f(x_{t-1}, u^{\bullet}_{t-1}, \hat{u}^{\circ}_{t-1}|_{\sigma^{\circ} = \varepsilon})
\end{equation}
where $f(\cdot, \cdot, \cdot)$ is the dynamics equation in \eqref{eqn:compact_dynamics_openloop}; $x_{t-1} = (x^{\bullet}, x^{\circ})$ is the observed game state at the previous time; $u^{\bullet}_{t-1}$ is the ego player's previous action; and $\hat{u}^{\circ}_{t-1}|_{\sigma^{\circ} = \varepsilon}$ is the predicted previous action of the other player assuming its role to be $\sigma^{\circ} = \varepsilon$, which is determined by the LFG model in~Section~\ref{sec:LFG_1}.

With a belief on the other player's role, we assume that the ego player adjusts its role probabilistically. 
In particular, we define a transition matrix to describe the probabilistic transition of the ego player's role:
\begin{equation}\label{eqn:role_transition_matrix_generic}
\begin{bmatrix} \pi^{\bullet}_{{\rm ll},t} & 1 - \pi^{\bullet}_{{\rm ff},t} \\
1 - \pi^{\bullet}_{{\rm ll},t} & \pi^{\bullet}_{{\rm ff},t} \end{bmatrix} = \Pi^{\bullet}_{t} = \Pi^{\bullet}_{t}(x_t, \bbp^{\rm p}(\sigma^{\bullet}_t)),
\end{equation}
where $\pi^{\bullet}_{{\rm ll},t}$ (resp. $\pi^{\bullet}_{{\rm ff},t}$) represents the probability of transitioning to role $\rm l$ (resp. $\rm f$) when the ego player's previous role is $\rm l$ (resp. $\rm f$); $\bbp^{\rm p}(\sigma^{\bullet}_t)$ represents a ``plausible'' distribution of the ego player's role, determined by the ego player's belief on the other player's role, i.e.,
\begin{equation}\label{eqn:plausiable_role}
    \bbp^{\rm p}(\sigma^{\bullet}_t) = \Phi\left(\bbp(\sigma^{\circ}| \xi_t)\right).
\end{equation}
The relation \UPDATE{$\Phi$} between the belief on the other player's role $\bbp(\sigma^{\circ}| \xi_t)$ and the ``plausible'' distribution of the ego player's role $\bbp^{\rm p}(\sigma^{\bullet}_t)$ can be designed. A reasonable design is
\begin{equation}\label{eqn:plausiable_role_example}
   \begin{bmatrix} \bbp^{\rm p}(\sigma^{\bullet}_t = {\rm l}) \\ \bbp^{\rm p}(\sigma^{\bullet}_t = {\rm f}) \end{bmatrix} = \begin{bmatrix} \bbp(\sigma^{\circ} = {\rm f}| \xi_t) \\ \bbp(\sigma^{\circ} = {\rm l}| \xi_t) \end{bmatrix}
\end{equation}
where ego tends to take a complementary role to the other player. 
We remark that alternative designs of $\Phi$ accounting for other considerations are also possible. 
Meanwhile, in \eqref{eqn:role_transition_matrix_generic}, $\Pi^{\bullet}_{t}(x_t, \bbp^{\rm p}(\sigma^{\bullet}_t))$ indicates that the transition probabilities can in general depend on the game state $x_t$ and the ``plausible'' distribution $\bbp^{\rm p}(\sigma^{\bullet}_t)$. For instance, when traffic state $x_t$ implies no conflict between two vehicles, the ego driver may ignore the ``plausible'' distribution $\bbp^{\rm p}(\sigma^{\bullet}_t)$ when determining his role.

\vspace{-1.0mm}
\section{Interactive Prediction and Planning Based on Leader-Follower Game with Adaptive Roles}\label{sec:interactive_planning}
\vspace{-1.0mm}
In this section, we present the interactive planning algorithm for the robot, assuming the human is making decisions based on an LFG model with adaptive roles. 
We first describe the method for getting interactive predictions, then present the planning algorithm.
\vspace{-2.0mm}
\subsection{Interactive Prediction}
\vspace{-2.0mm}
The robot assumes that the human behaves as an LFG model with role adaptation. 
Because the role of the human is not directly observable to the robot, the robot holds a belief on the role of the human and updates it based on Bayesian inference similarly to~\eqref{eqn:role_prop_update}.

Denote the belief at time $t$ as $\bbp(\sigma^{\rm H}_{t})$, then a corresponding open-loop prediction of human's behavior can be given by 
\begin{equation}
    \hat{x}^{\rm H}_{t+\tau} = x^{\rm H}_{\sigma^{\rm H}_{t}, t+\tau}, \quad \tau = 0, \ldots, T_1, \quad \sigma_{t}^{\rm H} \sim \bbp(\sigma^{\rm H}_{t})
\end{equation}
where it is assumed that the human will not change his role for $[t,t+T_1]$. 
At $t+T_1$, the human updates his belief on the robot's role $\bbp(\sigma_{t+T_1}^{\rm R})$, based on the game states resulting from the actions planned by the robot between $[t, t+T_1]$, determines a plausible distribution of his role $\bbp^{\rm p}(\sigma^{\rm H}_{t+T_1})$, and adjusts his role probabilistically, all using the methods in Section~\ref{sec:LFG_2}. 
Specifically, an adjusted role of the human is given as
\begin{equation}\label{eqn:role_update_at_T1_approx}
    \sigma_{t+T_1}^{\rm H} \sim \bbp(\sigma^{\rm H}_{t+T_1}) = 
    \hat{\Pi}(\hat{x}_{t+T_1}, \bbp^{\rm p}(\sigma_{t+T_1}^{\rm H}))\bbp(\sigma^{\rm H}_{t}).
\end{equation}
Note that the robot may not know the actual transition matrix of the human; thus, the ``$\wedge$'' in \eqref{eqn:role_update_at_T1_approx} indicates that the transition matrix used by the robot to predict the human role adjustment may be different from the human's actual transition matrix \eqref{eqn:role_transition_matrix_generic}.

With the adjusted role of the human, $\sigma_{t+T_1}^{\rm H}$, a corresponding prediction of human's behavior over the remaining horizon $[t+T_1,t+N]$ can be given as 
\begin{equation}
    \hat{x}^{\rm H}_{t+\tau} = x^{\rm H}_{\sigma^{\rm H}_{t+T_1}, t+\tau}, \,\,\, \tau = T_1 + 1, \ldots, N, \,\,\, \sigma_{t+T_1}^{\rm H} \sim \bbp(\sigma^{\rm H}_{t+T_1}).
\end{equation}
The predicted human actions over the horizon are given as
\begin{equation}
    \hat{u}^{\rm H}_{t+\tau} = 
    \begin{cases}
     \hat{u}_{\sigma_{t}, t+\tau}^{{\rm H}}, & \tau = 0, \ldots, T_1 - 1, \\
     \hat{u}_{\sigma_{t + T_{1}}, t+\tau}^{{\rm H}}, & \tau = T_1, \ldots, N -1.
    \end{cases}
\end{equation}
We note that due to the dependence on the human's adjusted role and hence his future behavior on the planned actions by the robot as shown in~\eqref{eqn:role_update_at_T1_approx}, the predictions are action-reaction interactive and closed-loop. That is, the predictions change as the robot's planned trajectory changes; see the interactive prediction scheme in~Fig.~\ref{fig:mpc_branch}. 

\vspace{-2.0mm}
\subsection{Interactive and Closed-Loop Planning}
\begin{figure}[t]
    \centering
    \includegraphics[width=0.9\linewidth]{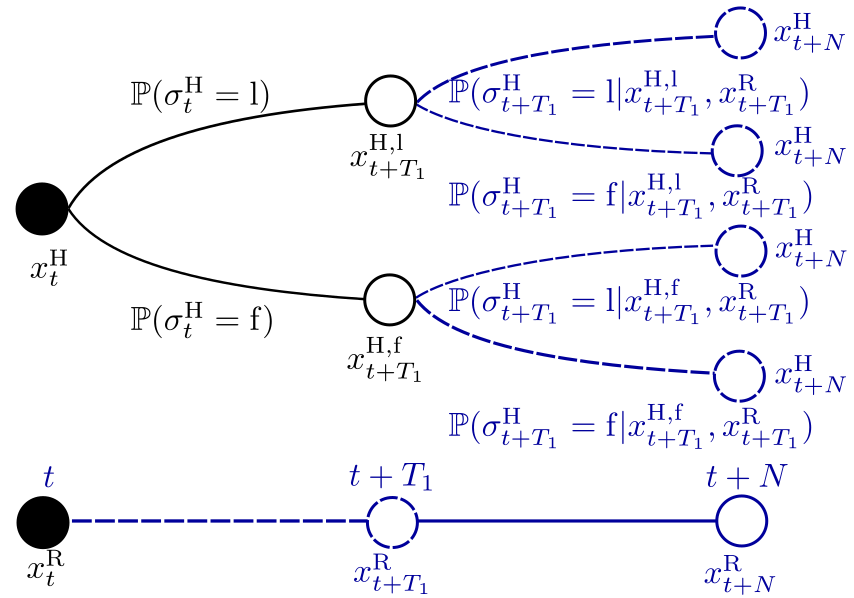}
    \caption{\small Closed-loop MPC-based strategy. The top branch corresponds to the prediction of human's behavior, while the lower branch corresponds to the robot's planned motion. 
    The black solid branches in the human prediction are affected by the state at $t$.
    The blue dashed branches are affected by the robot's behavior between $[t, t+T_{1}]$, which is also marked as a blue dashed branch, through the change of role $\sigma_{t+ T_{1}}^{\rm H}$. 
    }
    \label{fig:mpc_branch}
\end{figure}

\vspace{-2.0mm}
With interactive predictions of the human behavior, the robot's MPC policy~\eqref{eqn:mpc_policy} is updated to have two stages with four branches, as shown in Fig.~\ref{fig:mpc_branch}:
\begin{align}
        \uu^{\rm R, \ast}_{t} &\in \arg\max \, \nonumber \\
          &\left\{\underset{|\Upsilon|}{\bbe}\left[\sum_{\tau = 0}^{T_{1}} \lambda ^{\tau}R(x_{t + \tau + 1}, u_{t+\tau}^{\rm R}, \hat{u}_{\sigma_{t}, t+\tau }^{{\rm H}})\bigg|\xi_{t}\right]\right.  + \nonumber \\
        & ~~\left.\underset{|\Upsilon|^2}{\bbe}\left[\sum_{\tau = T_{1} + 1}^{N} \lambda ^{\tau} R(x_{t + \tau + 1}, 
        u_{t+\tau}^{\rm R}, 
        \hat{u}_{\sigma_{t + T_{1}}, t+\tau}^{{\rm H}})\bigg|\xi_{t}\right] \right\}  \nonumber \\
        \text{s.t. } \quad&  x_{t+\tau + 1} = f(x_{t+\tau}, \hat{u}^{\rm H}_{t+\tau}, u^{\rm R}_{t+\tau}),  \nonumber\\
    & \bbp(x_{t+\tau} \in \Xcal_{\rm safe}, \tau = 1, \ldots, N|\xi_{t}) \ge 1-\epsilon. \label{eqn:MPC_planning}
\end{align}
The expectations in \eqref{eqn:MPC_planning} are with respect to the probabilities $\bbp(\sigma_{t}^{\rm H})$ and $\bbp(\sigma_{t+ T_{1}}^{\rm H})$. 
The $|\Upsilon|$ denotes the number of roles. 
For each MPC decision made by the robot on the trajectories, there will be two possible futures, and the chance constraints will be evaluated uniformly for all states along the trajectory. 

Note that in the robot's prediction step, we assume that the initial role estimation by the human on the robot at $t$ is $[0.5, 0.5]$. 
In practice, the human could also use a Bayesian filter. 
This means that, in contrast to the robot generating predictions on the human's role continuously, the robot's perception of the human's estimation on the robot's role is not carried over between executions of \eqref{eqn:MPC_planning}.

\vspace{-1.0mm}
\section{Case Study: Autonomous Driving in
Interactive Traffic Scenarios}\label{sec:case_study}
\vspace{-1.0mm}

In this section, we apply the proposed closed-loop interactive planning to autonomous vehicles in the intersection where they need to interact with human-driven vehicles\footnote{\footnotesize Source codes and parameters: \url{https://github.com/CHELabUB/game_theoretic_model/tree/master/2025_mecc}}.

\vspace{-2mm}
\subsection{Scenario Setup}
\vspace{-2mm}
\subsubsection{Simulation Setup and Goals}
The top view of the simulation scenario is shown in Fig.~\ref{fig:frame_compare}, where
the autonomous vehicle (AV, red) is traveling eastbound, while the human-driven vehicle (HV, blue) is traveling northbound in a 2-way intersection. 
The merging point is at (0, 0) [m], and the white solid crossing line $6.5\, [\rm m]$ from the merging point indicates the entrance to the intersection.
Whoever reaches this crossing line first is considered to arrive first and will have the right of way at the intersection. 
Both the AV and HV aim to travel safely through the intersection in the fastest possible way. 
To achieve this, the AV could pass through the intersection faster than the HV by persuading the human driver to give it the right of way. 

We use the following discrete-time model to represent vehicle kinematics in a 2-way 1-lane intersection; c.f., \eqref{eqn:compact_dynamics_openloop}. 
\begin{equation}\label{eqn:vehicle_kinematics}
\begin{bmatrix}
   s^{\bullet, k}_{t+1} \\
   v^{\bullet,k}_{t+1}
\end{bmatrix} = 
\begin{bmatrix}
    1 & \Delta t \\ 0 & 1
\end{bmatrix}
\begin{bmatrix}
   s^{\bullet,k}_{t} \\
   v^{\bullet,k}_{t}
\end{bmatrix}
+ 
\begin{bmatrix}
   \frac{\Delta t^2}{2} \\
   \Delta t
\end{bmatrix}a_{t}^{\bullet,k},
\end{equation}
where $s, v, a$ denotes position, velocity, and acceleration, respectively. 
The subscript $t$ represents the discrete time, the first superscript $\bullet\in\{\rm H, R\}$ distinguishes the two players in the game, while the second superscript $k\in\{x, y\}$ denotes the $x$ or $y$-direction, $\Delta t$ is the sampling period. Since it is 1-lane, no lane change action is needed.

\begin{figure*}[!t]
    \centering
    \includegraphics[width=\linewidth]{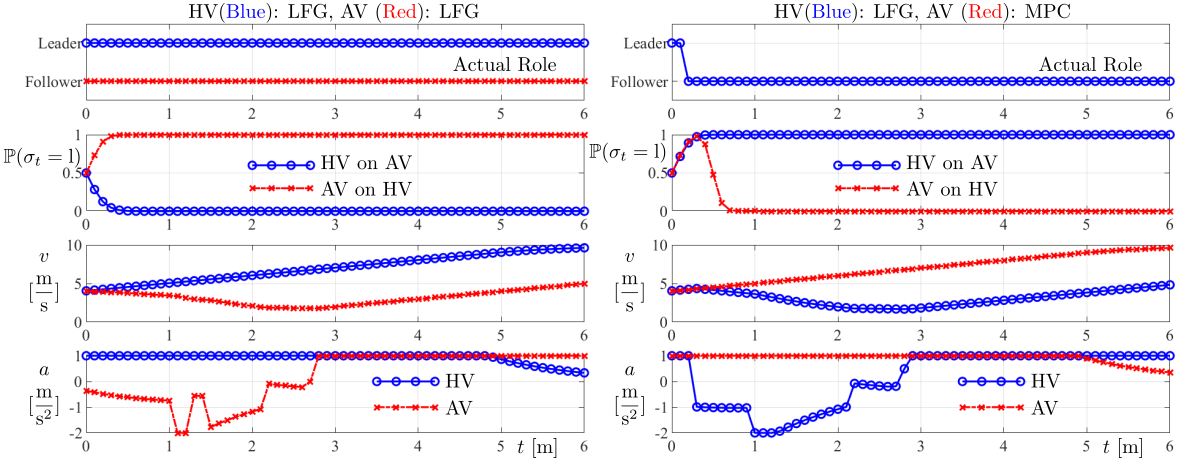}
    \caption{\small Time profile comparison plots. 
    The left column corresponds to the run when both human and AV are using the leader-follower game (LFG) to make decisions, while the right column corresponds to the run when human and AV are using LFG and MPC model \eqref{eqn:MPC_planning} to make a decision. In all panels, the blue profile corresponds to a blue human-driven vehicle that is traveling northbound, while the red profile corresponds to the red autonomous vehicle (AV) that is traveling eastbound. The first row shows the actual role of those who use the LFG model. The second row shows the estimated probability of holding a leader role that one agent holds on the other agent. The third row shows the speed profile, and the fourth row shows the acceleration profiles.}
    \label{fig:profile_compare}
\end{figure*}

\subsubsection{Safety Constraints}
To avoid collisions in planning, the safety requirements require staying within the safe set
\begin{equation}\label{eqn:collision_criteria}
    \Xcal_{\rm safe}  :=\left\{\left(s^1, s^2\right) \mid\left\|s^1-s^2\right\|_2 \geq 1.5 l_{\rm veh}\right\},
\end{equation}
where $s^{\bullet} = [s^{\bullet, x}, y^{\bullet, y}]^{\top}$, $\| \cdot \|_{2}$ represents Euclidean norm, and $l_{\rm veh} = 5$ [m] is the vehicle length. 
Given the uncertain nature of human prediction, \eqref{eqn:collision_criteria} is enforced as the chance constraint over the planning horizon in \eqref{eqn:MPC_planning}, with $\epsilon = 0.02$

\begin{equation}\label{eqn:collision_chance_constraint}
\bbp\left\{\left(s_{\tau \mid t}^1, s_{\tau \mid t}^2\right) \in \Xcal_{\rm safe}, \forall \tau=1, \cdots, N \mid \xi_t\right\} \geq 1 - \epsilon.  
\end{equation}
\vspace{-0.1mm}

\subsubsection{Reward}
The reward function is set to the following linear combination of terms, corresponding to liveliness (goal achievement), control effort, and collision \cite{he2024ecodriving}.
When the collision is inevitable, to avoid that vehicle establishing dangerous behavior by traveling at maximum speed, the collision penalty terms have speed-dependent coefficients \cite{li2020game}.
    \begin{align}
    \small
        c_{t} & = (1 + v_{t})r_{\rm collision}, \nonumber \\ 
        R_{t} & = s_{t}^{\bullet, k} - s_{0}^{\bullet, k} - w_{1} c_{t} - w_{2} |u_{t-1}|, \label{eqn:reward_intersection}
    \end{align}
where $(\bullet, k) = ({\rm R}, x)$ for the AV and $(\bullet, k) = ({\rm H}, y)$ for the HV. 
Horizon $N$ is set to 5 seconds to allow reaching and passing the intersection within the planning horizon from the beginning. 
Note that the indicator $r_{\rm collision} = 1 $ when a collision happens at $t$, and it is defined as the safety constraints \eqref{eqn:collision_criteria} being violated, i.e., $ s^{\bullet} \not\in \Xcal_{\rm safe}$. 
Note that \eqref{eqn:MPC_planning} also uses reward \eqref{eqn:reward_intersection} to deal with infeasible cases.

\subsubsection{Action Space}
 We carefully choose a set of trajectories as the action space by setting a list of target speeds and using an optimal velocity model to generate smooth and human-like trajectories that achieve different target speeds at designated spots (e.g., stop at the crossing line); see \cite{he2024ecodriving} for details. 
Given the acceleration range $a_{\max}$ and $a_{\min}$, the target speed over the horizon is selected to be $n_{\rm mesh}$ value between the maximum (capped at speed limit $v_{\max}$) and minimum reachable speed (capped at stop). 
Note that the trajectory associated with the minimum reachable speed as the target speed may be subject to changes for efficiency considerations: if the minimum reachable speed is 0, and it is possible to stop at the intersection, then the trajectory that leads to a stop at the intersection is used.
One additional trajectory that keeps the current speed is also added to $\UU_{\rm \bullet}$, making the total action trajectory of $|\UU_{\rm \bullet}|$ = $n_{\rm mesh} + 1$. In this case study, we set $n_{\rm mesh} = 10$.
In both the LFG model (\ref{eqn:follower_decision_strategy}, \ref{eqn:leader_decision_strategy}) and the MPC-based model \eqref{eqn:MPC_planning}, the trajectories are selected from this set of action trajectories, that is $\uu \in \UU_{\rm \bullet}$

\subsubsection{Role Transition Probability Matrix}
We assume the HV uses an LFG model with the plausible role distribution $\bbp^{\rm p}(\sigma_{t}^{\rm H})$ given by maximum likelihood estimation 
\begin{equation}
    \sigma^{\rm H}_{t} = 
    \begin{cases}
        {\rm l} & \text{if}\quad \bbp(\sigma_{t}^{\rm R} = {\rm l}) < \bbp(\sigma_{t}^{\rm R} = {\rm f}), \\
        {\rm f} & \text{if}\quad \bbp(\sigma_{t}^{\rm R} = {\rm l}) > \bbp(\sigma_{t}^{\rm R} = {\rm f}), \\
        \sigma_{t} & \text{if}\quad \bbp(\sigma_{t}^{\rm R} = {\rm l}) = \bbp(\sigma_{t}^{\rm R} = {\rm f}).
    \end{cases}
\end{equation}
Note that the last condition is a tie-breaker: the agent will keep the current role unchanged when the two probability values are identical.
Based on $\bbp^{\rm p}(\sigma_{t}^{\rm H})$, the role transition probability matrix is given by  
\begin{equation}\label{eqn:role_transition_matrix_intersection}
\Pi = \bbp^{\rm p}(\sigma_{t}^{\rm H} = {\rm l})
\begin{bmatrix}
    1 & p_{\rm a} \\
    0 & 1 - p_{\rm a}
\end{bmatrix}
+\bbp^{\rm p}(\sigma_{t}^{\rm H} = {\rm f})
        \begin{bmatrix}
             1 - p_{\rm a} & 0 \\
             p_{\rm a} & 1
        \end{bmatrix}
\end{equation}
cf.,\eqref{eqn:role_transition_matrix_generic}.
$p_{\rm a} \in [0, 1]$ captures the likelihood of the human to alter its role to the plausible role if different, which also capture how easy the human may be influenced. Note that with \eqref{eqn:role_transition_matrix_intersection}, human will keep the role if it is the same as the plausible role.
We remark that the main parameter $p_{\rm a}$ can be different from those used in \eqref{eqn:MPC_planning}, which is denoted as $\hat{p}_{\rm a}$. 
This corresponds to the fact that $\hat{\Pi}$ in \eqref{eqn:role_update_at_T1_approx} can be different from $\Pi$ in \eqref{eqn:role_transition_matrix_generic}, i.e., HV's $\Pi$ is unknown to AV. 

\vspace{-2.0mm}
\subsection{Results}
\vspace{-2.0mm}

\subsubsection{2-LFG Case}
We first simulate the case when both AV and HV use the LFG model.
The initial roles of HV and AV are complementary, and their estimation of the other vehicle starts from neutral. 
For both LFG models, we set $p_{\rm a} = 1$. 
The initial speed and position for the AV are fixed at $-20\,\rm[m]$ and $4\,[\rm m/s]$, while those for the human-driven vehicle are centered around $-20\,\rm[m]$ and $4\,[\rm m/s]$ with uniform distribution of $\pm 1[\rm m/s]$ and $\pm 5[\rm m]$.

\begin{table}[!h]
    \centering
    \begin{tabular}{|c|c|c|}
      \hline 
       Initial role by HV & AV [\%] & HV [\%]\\ \hline 
      Leader & 0.0 & 100.0   \\ \hline 
      Follower & 97.9 & 2.1  \\ \hline 
       \end{tabular}
    \caption{\small Percentage of arrival first at the crossing line based on 1000 runs with the random initial condition by the HV holding leader role vs follower role. 
    }
    \label{tab:2_game_stats}
    \vspace{-0.3cm}
\end{table}

We ran 1000 runs, and the results are summarized in Table \ref{tab:2_game_stats}. No collision or safety constraint violation occurs.
If the HV starts with a follower role, it will finish after the AV most of the time. 
By contrast, if it starts with a leader role, it will finish before the AV all the time. 
The bias arises from the fact that the initial condition is randomized for HV, whereas AV always starts from the same place at the same speed. 
One example of such runs is shown in the left column of Fig.~\ref{fig:profile_compare} for the time profile and Fig.~\ref{fig:frame_compare} for the top view lapse.
The role estimation by both players on the other converges to the correct role quickly, and neither agent changes their role. 
In these cases, the initial conditions and role (effectively fixed role) determine the outcome. 
These results show that while LFG produces causal and meaningful behaviors, the AV is unable to influence the HV to make persuasive planning decisions.

\begin{table}[!h]
    \centering
    \begin{tabular}{|c|c|c|c|}
      \hline 
      Initial Role by HV, $p_{\rm a} / \hat{p}_{\rm a}$ &AV [\%] & HV[\%] \\ \hline 
      Leader,\,1.00/1.00 & 87.8 & 12.2 \\ \hline 
      Follower,\,1.00/1.00 & 96.4 & 3.6\\ \hline 
      \hline 
      Leader,\,0.70/1.00 & 86.3 & 13.7 \\ \hline 
      Leader,\,0.50/1.00 & 82.9 & 17.1 \\ \hline 
      Leader,\,0.30/1.00 & 78.1 & 21.9 \\ \hline
      Follower,\,0.50/1.00 & 96.5 & 3.5  \\ \hline   
      Follower,\,0.70/1.00 & 96.5 & 3.5  \\ \hline
      Follower,\,0.30/1.00 & 96.6 & 3.4  \\ \hline 
      \hline 
      Leader,\,1.00/0.98 & 78.9 & 21.1  \\ \hline 
      Leader,\,1.00/0.95 & 0.0 & 100.0 \\ \hline 
      Leader,\,1.00/0.70 & 0.0 & 100.0  \\ \hline 
      Follower,\,1.00/0.98 & 96.4 & 3.6  \\ \hline 
      Follower,\,1.00/0.95 & 38.9 & 61.1 \\ \hline 
      Follower,\,1.00/0.70 & 1.2 & 98.8  \\ \hline 
       \end{tabular}
    \caption{\small Percentage of arrival first at the crossing line based on 1000 runs with the random initial condition by the human-driven vehicle. 
    $p_{\rm a} = \hat{p}_{\rm a}$ when AV knows HV's willingness to change the role;
    $p_{\rm a} > \hat{p}_{\rm a}$($p_{\rm a} < \hat{p}_{\rm a}$) when AV underestimates (overestimates) HV's willingness to change the roles.}
     \label{tab:mpc_game_stats}
     \vspace{-3.0mm}
\end{table}

\subsubsection{LFG + MPC Cases}
\begin{figure}
    \centering
    \includegraphics[width=0.98\linewidth]{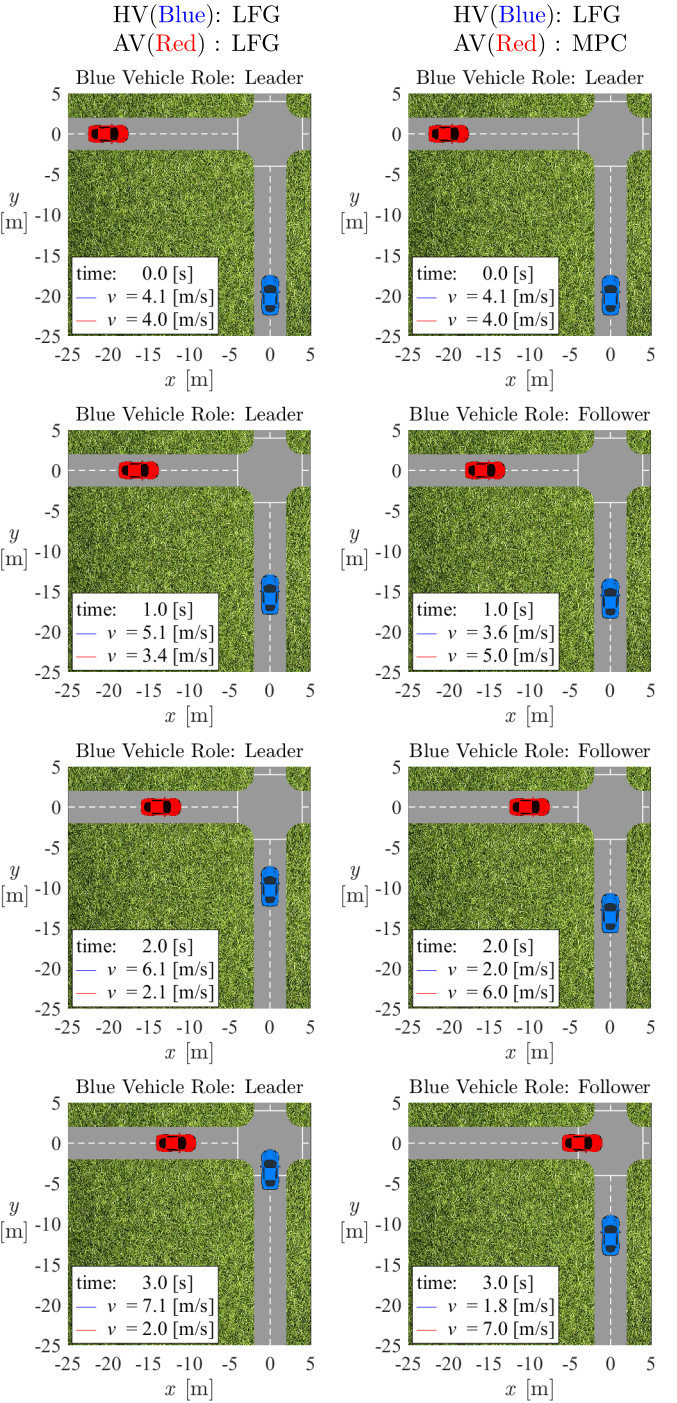}
    \caption{\small Topview lapse comparison. The left column corresponds to the run when both human and AV are using the leader-follower game (LFG) model to make decisions, while the right column corresponds to the run when human and AV are using LFG and MPC model \eqref{eqn:MPC_planning} to make a decision. In all frames, the blue vehicle and legend correspond to the HV, while the red vehicle corresponds to the AV, and their time profiles are shown as blue and red curves in Fig.~\ref{fig:profile_compare}.} 
    \label{fig:frame_compare}
\end{figure}

Next, we simulated the cases with the same setup except that the HV uses an LFG model while the AV uses the MPC algorithm \eqref{eqn:MPC_planning}.
We started the case when the parameter $p_{\rm a}$ is known by the AV, that is $p_{\rm a} = \hat{p}_{\rm a} = 1$, the prediction by the AV on the HV is accurate.
One example is shown in the right column of Fig.~\ref{fig:profile_compare} for the time profile and Fig.~\ref{fig:frame_compare} for the top view lapse.
Despite the HV starting with a leader role, and the AV holding a neutral estimation of the HV's role, the AV can persuade the HV to change roles, slow down, and allow the AV to cross the intersection first.
The statistics for this case are summarized in the first two rows of Table \ref{tab:mpc_game_stats}. 
Compared to the statistics in Table \ref{tab:2_game_stats}, for the majority of the cases, the AV can successfully persuade humans to give it priority.

To account for uncertainty, we then simulated cases where the parameter $p_{\rm a}$ varies at different levels unknown to AV, and report the results in Table \ref{tab:mpc_game_stats}. 
No collision or safety constraint violation occurs in any of the cases, demonstrating the safety of the design.
When the AV overestimates the HV's likelihood to adapt role ($\hat{p}_{\rm a} >p_{\rm a}$), it can still persuade the HV to change its role and give it the priorities, more likely when the HV starts with a follower role than a leader role, and can tolerate large gap between $\hat{p}_{\rm a}$ and $ p_{\rm a}$; See the second group in Table \ref{tab:mpc_game_stats}.
When the AV underestimates the HV's likelihood to adapt role ($\hat{p}_{\rm a} < p_{\rm a}$), the AV with MPC algorithm would choose not to persuade HV and instead give HV more priority as the gap between $\hat{p}_{\rm a}$ and $ p_{\rm a}$ increases; See the third group in Table \ref{tab:mpc_game_stats}. 
We acknowledge that the big swing from AV finish first mostly to HV finish first mostly between $\hat{p}_{\rm a} = 0.98$ and $0.95$ is associated with the chance constraint level set $\epsilon =0.02$; cf. \eqref{eqn:collision_chance_constraint}.

In summary, bearing uncertainties in the key parameters, the proposed MPC-based interactive planning algorithm \eqref{eqn:MPC_planning} enables the AV to be persuasive, influence human behavior, and gain an advantage.

\vspace{-1.0mm}
\section{Conclusion} \label{sec:conclusion}
\vspace{-1.0mm}
In this study, we presented a framework that enables autonomous vehicles to proactively shape the intentions and behaviors of interacting human drivers. 
The framework employed a leader-follower game model with an adaptive role mechanism to predict human interaction intention and behavior, and utilized a branch model predictive control algorithm to plan the AV trajectory that persuades the human to adopt a desired intention. 
We demonstrated the proposed framework in an intersection scenario, and simulation results illustrated its effectiveness in generating persuasive AV trajectories under uncertainties.
Future directions include implementing and validating the proposed framework in more complex driving scenarios, \UPDATE{with human-in-the-loop, and baseline against other non-persuasive motion planning algorithms.}

\vspace{-3.0mm}
\bibliography{mecc2025}

\end{document}